\documentclass[prl,twocolumn,showpacs,floatfix,amsmath,amssymb]{revtex4}
\usepackage{graphicx}
\usepackage{dcolumn}
\usepackage{bm}
\usepackage{float}
\usepackage{mathrsfs}
\usepackage{SIunits}
\newcommand{\dl}{\partial}

\begin{document}
\title{Phonon instability and self-organized structures in multi-layer stacks of confined dipolar Bose-Einstein condensates in optical lattices}
\author{Patrick K\"oberle}
\email{koeberle@itp1.uni-stuttgart.de}
\author{G\"unter Wunner}
\affiliation{Institut f\"ur Theoretische Physik 1, Universit\"at Stuttgart, 70550 Stuttgart, Germany}
\date{\today}

\begin{abstract}
In calculations to date \cite{wang08,klawunn09}  of multi-layer stacks of dipolar condensates, created in one-dimensional optical lattices, the condensates have been assumed to be two-dimensional. In a real experiment, however,
the condensates do not extend to infinity in the oblate direction, but have to be confined by a trap potential, too. By three-dimensional numerical simulations of this realistic experimental situation we find a crucial dependence of the phonon instability boundary on the number of layers.  Moreover, near the boundary of the phonon instability, a variety of structured ground-state wave functions emerges, which may indicate the onset of a roton instability \cite{ronen07, wilson08}.

\end{abstract}


\maketitle

\section{Introduction}

The experimental realization of Bose-Einstein condensates 
of chromium atoms \cite{griesmaier05}, with a strong interatomic magnetic dipole-dipole interaction, has
opened the door to promising new experiments  \cite{stuhler05} on dipolar gases in which
a wealth of new phenomena predicted by theory, such as the appearance of radial and angular rotons, 
biconcave shapes of the ground state, anisotropic solitons, etc. 
\cite{giovanazzi02,santos03,li00,li04,dell04,Ron06,Ron07,dutta07,Tik08}, should be observable. In dipolar condensates, 
the manipulation of the $s$-wave scattering length via Feshbach resonances offers
the opportunity of tuning the relative strength of the interactions
through the whole range,  from dominance of the short-range interaction 
to that of the dipole-dipole interaction, and of studying the condensates in the
different regimes.  

Recently, stacks of non-overlapping dipolar
condensates, unconfined in two dimensions, in one-dimensional optical lattices have been investigated theoretically  \cite{wang08,klawunn09}. Special attention has been payed to the well-known roton-instability, and it has been reported in Ref. \cite{klawunn09} that this instability is enhanced, in the sense that the condensates become dynamically unstable in a wider range of the $s$-wave scattering length (measured in units of the dipole length) when more layers are added to the stack. Faraday pattern size
transitions have been proposed \cite{nath09} as an experimental probe for
the onset of the roton minimum in multi-layer stacks of two-dimensional
dipolar condensates. 
The phonon instability, however, in those calculations always set in at the same value of the ratio of contact and dipole-dipole interaction, independent of the number of layers. 

In an experiment a confining trap potential
inevitably has to be present also in the planes of the condensates. It is the
purpose of this paper to carry out three-dimensional numerical simulations
for this realistic situation, investigating the \textit{phonon} instability in more detail. It will turn out that the latter
depends crucially on the stack size. A second consequence of the three-dimensional confinement is the appearance of a variety of structured ground states in the vicinity of the phonon instability. It has been suggested in the literature \cite{ronen07, wilson08} that the emergence of structures in the condensate density is related to the onset of a roton instability. 
\begin{figure}
\includegraphics[width=0.5\columnwidth]{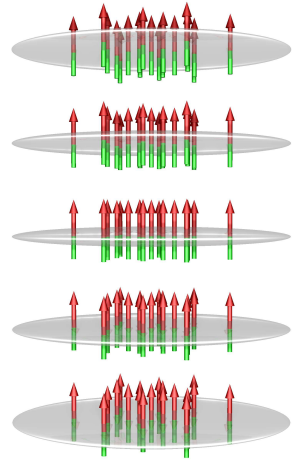}
\caption{\label{stack}
A stack of $N_s$ dipolar Bose-Einstein condensates. The long-range nature of the dipole-dipole interaction couples all individual condensates. The traps are assumed to be extremely oblate and inter-site hopping of particles is prohibited.}
\end{figure}
\section{Theory}
As in Ref. \cite{klawunn09} we  consider a stack of $N_s$ dipolar Bose-Einstein condensates, each 
containing the same number of atoms $N$.  We consider axisymmetric traps, whose geometry is  specified \cite{koch07} by the
 mean trap frequency $\overline{\omega} = (\omega_{\varrho}^2 \omega_z)^{1/3}$ and the aspect ratio $\lambda = \omega_z / \omega_{\varrho}$ ($\omega_{\varrho, z}$ are the trap frequencies in the $\varrho$- and $z$-direction, respectively).
The spacing between adjacent traps, $\Delta$, must be large enough for condensate wave functions  not to overlap (i.e. inter-site hopping of atoms is forbidden), but sufficiently small for neighbouring condensates to "feel" each other, despite the relative weakness of the dipolar interaction. Hence the traps are extremely oblate (aspect ratios $\lambda > 100$).  Fig. \ref{stack} shows a schematic representation of the system.\\
As is well known, at sufficiently low temperatures the ground state of a single condensate of weakly interacting bosons can be represented by a single wave function, whose dynamics obeys the Gross-Pitaevskii (GP) equation. For a stack of condensates the wave functions in different layers are described by a {\em set} of coupled extended GP equations. To obtain this set of equations in dimensionless
form, we introduce  "atomic" units \cite{koeberle09}, derived from  the mass $m$ and the magnetic dipole moment $\mu$ of an atom,
namely a "dipole length''
$	a_d^{} = {m \mu_0 \mu^2}/(2 \pi \hbar^2)$,
a unit energy $E_d^{} = \hbar^2/(2 m a_d^{2})$, a unit frequency $\omega_d^{} = E_d^{} / \hbar$ and a unit time $t_d^{} = \hbar / E_d^{}$, respectively. In addition, to eliminate the explicit dependence of the 
interaction potentials  on the 
number of atoms, we carry out a particle number scaling for the position and time coordinates,
$	\mathbf{r} \rightarrow N a_d^{} \mathbf{r} \; , 
	t \rightarrow N^2 t_d^{} t.$
In the scaled atomic units, the coupled system of GP equations for the set of $N_s$ condensate wave functions
$\{\psi_j(\mathbf{r}, t)\}$ then reads:
\begin{widetext}
\begin{equation}\label{GP_trapped}
	\left( \hat{H}_{0j} + 8 \pi a \left\vert \psi_j (\mathbf{r}, t) \right\vert ^2 + \sum_{k=1}^{N_s}  \int d r'^3 \, \frac{1-3 \cos^2 \vartheta '}{\left\vert \mathbf{r} - \mathbf{r}' \right\vert^3} \left\vert \psi_k (\mathbf{r}', t) \right\vert^2 \right) \psi_j (\mathbf{r}, t) = i \dl _t \psi_j (\mathbf{r}, t), \quad\quad (j = 1, \dots, N_s)\,.
\end{equation}
\end{widetext}
The kinetic energy and the trap potential terms for the condensate in layer $j$ have been subsumed in  
$\hat{H}_{0j}$,
\begin{equation}\label{H0j}
	\hat{H}_{0j} = -\Delta_{\mathbf{r}} + N^4 \gamma_{\varrho}^2 \rho^2 + N^4 \gamma_z^2 \left(z + \frac{N_s+1-2j}{2} \Delta \right)^2\,,
\end{equation} 
where $ \gamma_{\varrho,z}= \omega_{\varrho,z}/(2\omega_d)$ denote the dimensionless trap frequencies.
It is evident from (\ref{GP_trapped}) and (\ref{H0j}) that the three external parameters governing
the system are  the scattering length $a$, measured in units of the dipole length $a_d^{}$, and the atom 
number scaled trap frequencies $N^2 \gamma_{\varrho,z}$, or, alternatively, the aspect ratio $\lambda=
\gamma_{\varrho}/\gamma_{z}$
and the atom number scaled mean trap frequency $N^2\overline{\gamma} = N^2 \gamma_{\varrho}^{2/3}\gamma_{z}^{1/3}$. We note that our parameter $N^2\gamma_{\varrho}$ is related to the dimensionless parameter $D$ 
used in Refs. \cite{dutta07,ronen07} to measure the effective strength of the dipole interaction in
trap units  by $D= (N^2\gamma_{\varrho}/4)^{1/2}$. We also note that the dimensionless parameter $\beta$
used by Klawunn and Santos \cite{klawunn09} to characterize the strength of the dipole interaction
compared to the short-range interaction is related to our dimensionless scattering length $a$ by
$\beta= 1/(3a)$.

Following Ref.  \cite{klawunn09},  we make use of the very large aspect ratio $\lambda$ which leads to a decoupling of the motions in the 
longitudinal and the transverse direction
\begin{equation}
	\psi_j(\mathbf{r}, t) = \psi_{\bot, j} (\varrho, t) \phi_{j} (z, t),
\end{equation}
and assume that in the longitudinal direction the condensate wave function is  frozen in the ground state of the 
harmonic $z$-trapping potential in layer $j$  
\begin{equation}
	\phi_{j} (z, t) = \pi^{-1/4}l_z^{-1/2}{e^{-\left(z+\frac{N_s+1-2j}{2} \Delta \right)^2/2l_z^2}} e^{-i \mu_z t},
\end{equation}
where
	$\mu_z =N^2 \gamma_z$ is the chemical potential of the ground state and $l_z = \sqrt{1/\mu_z}$
the oscillator length.
Multiplying the GP equation (\ref{GP_trapped}) by $\phi_{j}$ and integrating over the $z$-coordinate yields
after a lengthy calculation\begin{widetext}
\begin{equation}\label{GP_trapped_adiabatic}
	\left( \hat{H}_0 + \frac{8 \pi a }{l_z \sqrt{2 \pi}} \left\vert \psi_{\bot, j} (\varrho, t) \right\vert ^2 + \sum_{k=1}^{N_s} \mathscr{H}^{-1} \left\lbrace \mathscr{H} \left\lbrace \left\vert \psi_{\bot, k} (\varrho, t) \right\vert ^2 \right\rbrace F(k_{\varrho}, j-k) \right\rbrace \right) \psi_{\bot, j} (\varrho, t) = i \dl_t \psi_{\bot, j} (\varrho, t),
\end{equation}
\end{widetext}
with
\begin{equation}
	\hat{H}_0 = - \Delta_{\varrho} + N^4 \gamma_{\varrho}^2 \varrho^2,
\end{equation}
and
\begin{equation}
	F(k_{\varrho}, n) = \frac{2}{3} \int\limits_{-\infty}^{\infty} d k_z \, \left( \frac{3 k_z^2}{k_{\varrho}^2 + k_z^2} - 1 \right) e^{-\frac{1}{2} k_z^2 l_z^2} \cos(k_z n \Delta).
\end{equation}
In (\ref{GP_trapped_adiabatic}), $\mathscr{H}$ and $\mathscr{H}^{-1}$ denote the Hankel transform and its inverse, respectively, which appear as a consequence
of the 2D Fourier transform in cylinder coordinates used to evaluate the convolution integral. The ground state is obtained by solving (\ref{GP_trapped_adiabatic}) in imaginary time on a discretized spatial grid. In the calculations 
we make use of the discrete Hankel transform method presented in \cite{ronen06}. 
\section{Results and discussion}
\begin{figure}
\includegraphics[width=0.7\columnwidth]{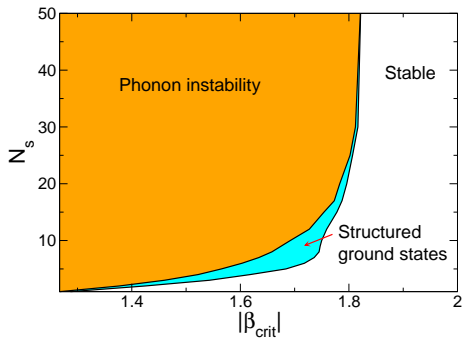}
\caption{\label{a_crit2}
Critical scattering length for the onset of roton- and phonon-instability depending on the number of layers. The mean trap frequency is $N^2 \bar{\gamma} = 1.3 \cdot 10^3$.}
\end{figure}
We simulate an experiment of a dipolar condensate of $^{52} \mathrm{Cr}$ atoms, for which $\mu = 6 \mu_{B}^{}$ and $a_d^{} = 91 a_{B}^{}$, with $a_{B}^{}$ and $\mu_{B}^{}$ the Bohr radius and the Bohr magneton, respectively. Fig. \ref{a_crit2} shows the results for the stability diagram for a mean trap frequency $N^2 \bar{\gamma} = 1.3 \cdot 10^3$, a trap aspect ratio $\lambda = 340$ and a spacing $\Delta = 0.035$. Choosing $\bar{\omega} = 2 \pi \cdot \unit{1.0}{\kilo \hertz}$, this implies approximately $3300$ particles per layer and a spacing of $\unit{0.56}{\micro\meter}$. These are typical numbers for actual experiments. It is evident 
that there appears a strong  dependence of the critical value of $\beta$ (or, in other words, the scattering length $a$, with $a<0$) on the number of layers. This instability threshold is denoted phonon instability due to the fact that excitations of the condensates with small or vanishing wave vector (phonons) lead to imaginary energy eigenvalues and consequently collapse of the condensates. The fact that the phonon instability moves to larger absolute values of $\beta$ (i.e. smaller absolute values of the scattering length)
when the number of condensates is increased can be understood in a very simple picture: With growing number of sites, the total dipolar potential to which the individual condensates are exposed becomes stronger. This potential forces the condensates to shrink radially which leads to a higher density, necessary for inducing collapse. Consequently, the collapse threshold is being pushed to larger absolute values of $\beta$. As our calculations show, it is the condensate in the middle of the stack which becomes unstable first. This is possibly not surprising, since the dipolar potential approaches a maximum in the middle layer. The phonon instability must therefore be understood in the sense that below the threshold, there is at least one layer, and not necessarily the whole system,  which becomes unstable.

At $N_s \approx 50$, $\beta_{\mathrm{crit}}$ approaches a constant value. Again, we can explain this by considering the condensate in the middle. The dipole-dipole interaction decays as $1/r^3$, and the sum over the dipolar potentials of all condensates converges with growing number of sites, as well as the phonon instability. We point out that the effect described can only appear in a 3D trap. In \cite{klawunn09}, the homogeneous case was considered, where the condensate density was the same for all layers and for any number of sites. As a consequence, the phonon instability appeared at the same value of $\beta$ independent of the number of sites.

\begin{figure}
\includegraphics[width=0.7\columnwidth]{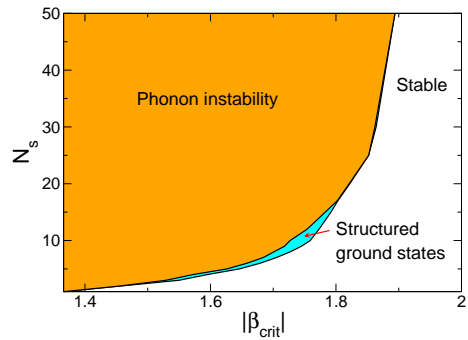}
\caption{\label{a_crit3}
Critical scattering length for the onset of roton- and phonon-instability depending on the number of layers. The mean trap frequency is $N^2 \bar{\gamma} = 1.3 \cdot 10^4$.}
\end{figure}

 Fig. \ref{a_crit3} shows the results for the stability diagram for a mean trap frequency of $N^2 \bar{\gamma} = 1.3 \cdot 10^4$ and a value of the spacing parameter of $\Delta = 0.011$. The latter value is reduced by a factor of $\sqrt{10}$ because we effectively increased the particle number in the system by a factor of $\sqrt{10}$. According to the scaling laws, in the experiment proposed, the spacing is still $\unit{0.56}{\micro\meter}$. As we can see in Fig. \ref{a_crit3}, in the limit $N_s \rightarrow \infty$, $\beta_{\mathrm{crit}}$ approaches a larger value than in Fig. \ref{a_crit2}. This is again due to the higher density in the condensates.

Another feature that already appears in single condensates is the formation of structures in the condensate wave function. It has been reported \cite{ronen07,wilson08} that near the phonon instability and at special trap aspect ratios, the shape of the condensate wave function differs from a "normal" Gaussian or parabola, where the maximum density lies at the centre of the trap, and instead forms a density pattern with the peak density away from the centre. Due to the special shape these density patterns
have been dubbed  as "blood-cell-like" condensates. 

In extremely oblate and layered condensates, the situation is qualitatively different.
\begin{figure}
\includegraphics[width=0.65\columnwidth]{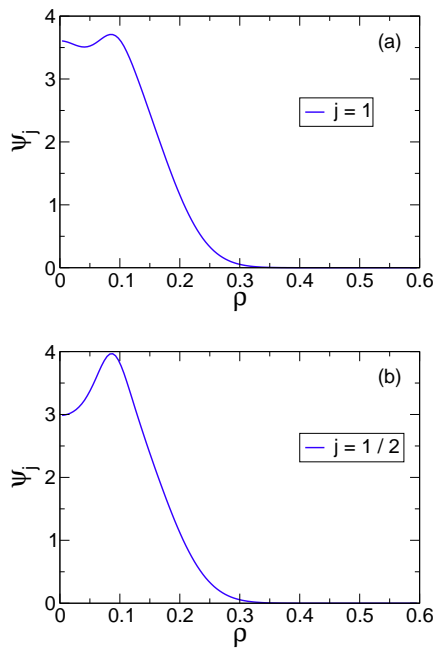}
\caption{\label{psi1}
Structures of the wave functions in the very vicinity of the phonon instability for a single trap (a) and 2 traps (b). The mean trap frequency is $N^2 \bar{\gamma} = 1.3 \cdot 10^3$.}
\end{figure}
\begin{figure}
\includegraphics[width=0.65\columnwidth]{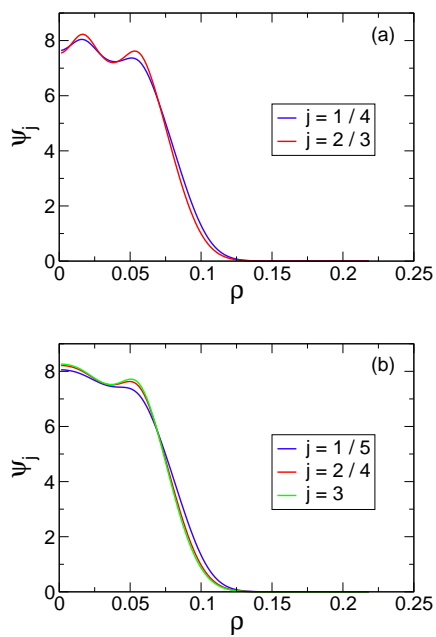}
\caption{\label{psi2}
Structures of the wave functions in the very vicinity of the phonon instability for 4 traps (a) and 5 traps (b). The mean trap frequency is $N^2 \bar{\gamma} = 1.3 \cdot 10^4$.}
\end{figure}
Fig. \ref{psi1}a shows the radial part of the wave function for a single condensate trapped at $N^2 \bar{\gamma} = 1.3 \cdot 10^3$ in the very vicinity of the phonon instability. One peak of the density distribution lies at the centre of the trap, but there is also a second, larger peak at $\varrho \approx 0.09$.  This picture changes when a second layer is added to the stack. The radial wave functions for the condensates $1$ and $2$ (which are identical due to the symmetry) now assume the familiar blood-cell shape with one density peak away from the centre of the trap. This structure remains when further layers are added. In the stability diagrams, Figs. \ref{a_crit2} and \ref{a_crit3}, the regions where these structures arise are coloured blue. When increasing the trapping frequency to $N^2 \bar{\gamma} = 1.3 \cdot 10^4$, the situation is even richer. Fig. \ref{psi2}a shows the radial wave functions for a stack of $4$ condensates. In all layers, the condensate wave functions exhibit $2$ peaks away from the centre, while the peaks are more pronounced for the two in-most condensates; this shows again the effect of the dipole-dipole interaction being enhanced by the outer condensates. In Fig. \ref{psi2}b, a fifth layer was added, and it now seems energetically more favourable for the condensates to move one peak to the centre, while keeping an outer, smaller peak.
\section{Conclusions}
 By three-dimensional numerical simulations of  multi-layer stacks of confined dipolar Bose-Einstein condensates in optical lattices
for realistic experimental parameters we have revealed a crucial dependence of the phonon instability boundary on the number of layers. This finding is at variance with the results of Ref. \cite{klawunn09} where unconfined 2D condensates were considered and the
critical value of the boundary was independent of the number of layers. We have also demonstrated that near the boundary of the phonon instability, a variety of structured ground-state wave functions emerges.

Structured ground-state wave functions have been intensively discussed in the literature \cite{ronen07, wilson08}. It has been shown that the emergence of structures is intimately related to the appearance of imaginary energy eigenvalues in the Bogoliubov spectrum. More precisely, the Bogoliubov modes that become unstable have non-vanishing projection of the angular momentum on the $z$-axis \cite{wilson08}. These angular roton modes are responsible for a dynamical instability of the condensates resulting in local collapse. Contrary to global collapse, where the particles are accumulated at the centre of the trap, a locally collapsing condensate forms several density peaks ordered on a ring. It is the same ring on which the ground-state wave function reaches a maximum. In our case, the situation is different since more than one density peak occurs. Evidently a 3D numerical investigation of the Bogoliubov spectrum along these lines in the region of structured ground states in Figs. \ref{a_crit2} and \ref{a_crit3} is urgent, and under way.

\begin{acknowledgments}
We thank Jonas Metz and Stefan M\"uller from 5. Physikalisches Institut, Universit\"at Stuttgart, for fruitful discussions.
\end{acknowledgments}

\bibliographystyle{apsrev}
\bibliography{bec_lattice,bec_bank_1.bib}

\end{document}